\documentclass[aps,prl,10pt,twocolumn,longbibliography,noeprint,floatfix,superscriptaddress]{revtex4-2}

\usepackage{amsmath, amssymb}
\usepackage{mathtools}
\usepackage{lipsum}
\usepackage{times}
\usepackage{graphicx}
\usepackage{wasysym}
\usepackage{bm}
\usepackage{hyperref}
\usepackage{xspace}
\usepackage{siunitx}
\usepackage[dvipsnames]{xcolor}
\usepackage{soul}
\usepackage{centernot}
\usepackage{braket}
\usepackage{microtype}
\usepackage{upgreek}
\usepackage{booktabs}
\usepackage{float}
\usepackage{capt-of}

\definecolor{customcolor}{rgb}{0.261, 0.212, 0.658}

\definecolor{myColor2}{rgb}{0.02,0.12,0.3}  

\definecolor{myColor}{rgb}{0.02,0.12,0.7}
\definecolor{myciteColor}{rgb}{0.39,0.7,0.89}
\hypersetup{colorlinks=true, linkcolor=myColor, filecolor=myColor, urlcolor=myColor, citecolor=myColor, urlcolor=myColor}
\DeclareSIUnit{\gauss}{\ensuremath{\mathrm{G}}}
\DeclareSIUnit{\bohrradius}{\ensuremath{a_0}}
\graphicspath{{./Figures/}}

\DeclareSIUnit{\nK}{\nano\kelvin}
\DeclareSIUnit{\um}{\micro\metre}
\DeclareSIUnit{\aB}{\emph{a}_0}
\DeclareSIUnit{\G}{G}

\newcommand{\appropto}{\mathrel{\vcenter{
  \offinterlineskip\halign{\hfil$##$\cr
    \propto\cr\noalign{\kern2pt}\sim\cr\noalign{\kern-2pt}}}}}

\def\be{\begin{equation}}
\def\ee{\end{equation}}

\makeatletter
\def\@fnsymbol#1{\ensuremath{\ifcase#1\or *\or \dagger\or \ddagger\or
   \mathsection\or \mathparagraph\or \|\or **\or \dagger\dagger
   \or \ddagger\ddagger \else\@ctrerr\fi}}
\makeatother
\newcommand{\um}{\upmu{\rm m}}
\newcommand{\nK}{\textrm{nK}}

\newcommand{\potassium}{^{39}\textrm{K}}

\begin{document} 
 
\title{
Local velocimetry for atomic superfluids: from Riemann rarefaction waves to quantized vortices
}

\author{Yansheng~Zhang}\email{yz661@cam.ac.uk}
\affiliation{Cavendish Laboratory, University of Cambridge, J. J. Thomson Avenue, Cambridge CB3 0US, UK}

\author{{Yi~Jiang}}
\affiliation{Cavendish Laboratory, University of Cambridge, J. J. Thomson Avenue, Cambridge CB3 0US, UK}

\author{Feiyang~Wang}
\affiliation{Cavendish Laboratory, University of Cambridge, J. J. Thomson Avenue, Cambridge CB3 0US, UK}

\author{Louis Depaepe}
\affiliation{Cavendish Laboratory, University of Cambridge, J. J. Thomson Avenue, Cambridge CB3 0US, UK}

\author{J.~Q. Matthew Chia}
\affiliation{Cavendish Laboratory, University of Cambridge, J. J. Thomson Avenue, Cambridge CB3 0US, UK}

\author{Christoph~Eigen}
\affiliation{Cavendish Laboratory, University of Cambridge, J. J. Thomson Avenue, Cambridge CB3 0US, UK}
\author{{Gehrig~Carlse}} 
\affiliation{Cavendish Laboratory, University of Cambridge, J. J. Thomson Avenue, Cambridge CB3 0US, UK}
\author{Zoran~Hadzibabic}
\affiliation{Cavendish Laboratory, University of Cambridge, J. J. Thomson Avenue, Cambridge CB3 0US, UK}
\begin{abstract}

We demonstrate local velocimetry for atomic superfluids in two dimensions (2D). 
Using matter-wave interferometry based on Raman--Bragg beam splitters, we map the local gradient of the phase of the superfluid's wavefunction onto displacement of the interference fringes and reconstruct the 2D velocity field.
We illustrate the capabilities of this technique by applying it to two canonical problems. First, we directly observe the self-similar evolution of the velocity field in 1D Riemann rarefaction flows.
Second, we map out the full 2D velocity field around quantized vortices, explicitly demonstrating the quantization of velocity circulation.
\end{abstract}
\maketitle

Atomic Bose--Einstein condensates (BECs) are prototypical superfluids~\cite{Pitaevskii:2016}, displaying dissipationless flow~\cite{Raman:1999, Ryu:2007, Desbuquois:2012} and quantized vortices~\cite{Matthews:1999, Madison:2000a, AboShaeer:2001}.
In different geometries, and with tunable microscopic parameters, they provide a versatile platform for studies of quantum-fluid phenomena, from solitary waves~\cite{Burger:1999,Denschlag:2000,Strecker:2002, Khaykovich:2002,Cornish:2006, Hamner:2011, Chen:2020, Bakkali-Hassani:2021} to shock waves~\cite{Donley:2001,Dutton:2001,Hoefer:2006,Chang:2008,Mossman:2018} and turbulence~\cite{Henn:2009, Neely:2013, Kwon:2014, Navon:2016, Gauthier:2019, Johnstone:2019, Navon:2019,Liu:2021, Reeves:2022, Galka:2022, Dogra:2023, Karailiev:2024, Gazo:2025, Zhao:2025, Simjanovski:2025, Jiang:2026, Morris:2026, Fischer:2026, Bradley:2026}.
For many such studies, the key properties of a superfluid, just like for a classical one, are its density and velocity fields.

\begin{figure}[!b]
    \centering
    \includegraphics[width=\linewidth]{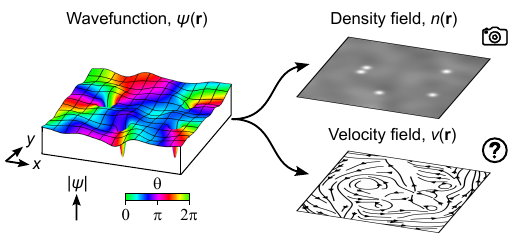}
    \caption{
    Density and velocity of an atomic superfluid (cartoon). The superfluid, here in 2D, is described by a complex wavefunction $\psi(\mathbf{r})$. Its density, $n(\mathbf{r}) = |\psi(\mathbf{r})|^2$, can be directly imaged with high resolution. In this illustration, the holes in the density distribution reveal the presence of vortices, but not the sign or magnitude of the associated velocity circulation.
    Generally, measuring the full velocity field, $\mathbf{v}(\mathbf{r)}\propto\nabla\theta(\mathbf{r})$, where $\theta (\mathbf{r})$ is the phase of $\psi(\mathbf{r})$, is challenging. 
    }
    \label{fig:1}
\end{figure}

For an atomic superfluid described by a complex macroscopic wavefunction $\psi(\mathbf{r})$ (Fig.~\ref{fig:1}), the density field, $n(\mathbf{r}) =|\psi(\mathbf{r})|^2$, can be directly imaged, with various microscopy techniques~\cite{Bakr:2009, Hung:2014, Asteria:2021,Xiang:2025,Yao:2025,deJongh:2025} providing high resolution images. However, measuring the velocity field,
\begin{equation}
    \mathbf{v}(\mathbf{r}) ={\hbar}/{m}\cdot\nabla\theta(\mathbf{r})\,,
    \label{eq:v-def}
\end{equation} 
where $m$ is the atom mass and $\theta(\mathbf{r})$ the phase of $\psi(\mathbf{r})$, is more challenging.

There are many techniques for extracting information related to the phase of an atomic gas. For example, the global momentum distribution of the system can be measured using
time-of-flight expansion~\cite{Anderson:1995, Davis:1995a}, momentum-focusing~\cite{Tung:2010}, or Bragg spectroscopy~\cite{Kozuma:1999a, Stenger:1999b}. Additionally, phase correlations can be studied using Talbot-style interferometry~\cite{Dettmer:2001, Choi:2012, Sunami:2025, Brueggenjuergen:2026, Su:2026}, matter-wave interference of two statistically equivalent clouds~\cite{Hadzibabic:2006, Hofferberth:2007, Sunami:2023}, or the interferometric contrast in the interference of a cloud and a translated copy of itself~\cite{Hagley:1999b,  Hugbart:2005, Clade:2009, Navon:2015,Rabga:2024}. These techniques probe phase coherence properties and, in principle, reveal information about the system's velocity distribution. However, they do not reveal a spatially resolved velocity field.

The local phase $\theta(\mathbf{r})$ can be revealed by interfering the cloud with a condensate that has a known phase profile. Experiments in this spirit have demonstrated, for example, the phase accumulated during the interaction-driven expansion of a BEC~\cite{Simsarian:2000}, the phase singularities associated with quantized vortices~\cite{Inouye:2001, Chevy:2001}, and the circulating currents in 1D ring geometry~\cite{Moulder:2012, Corman:2014,Eckel:2014, Mkrtchyan:2026}; see also End Matter for an example specific to our 2D system. From the measured $\theta(\mathbf{r})$, one can in principle, by numerical differentiation, obtain the velocity field. However, a suitable phase reference needed for this approach to velocimetry is in general not available.

A number of experiments have also explored approaches closer to how $\mathbf{v}(\mathbf{r})$ is measured in classical fluids. Particle tracing~\cite{Zhao:2025, Sharan:2025}, time-resolved density measurements combined with the continuity equation~\cite{Lahav:2010, Murtadho:2025}, and the Doppler effect on Bragg scattering~\cite{Muniz:2006, Seo:2017, Johnstone:2019} have all been used to partially reconstruct the local velocity fields. 
Despite these developments and further proposals~\cite{Murthy:2019, Cooper:2026, YYang:2026} (see also Refs.~\cite{Impertro:2024, Brueggenjuergen:2026, Weitenberg:2026, Su:2026} for lattice systems and Refs.~\cite{SMoore:2016,Sulpizio:2019, XLiu:2021, Aharon-Steinberg:2022, Panico:2023, Svancara:2024} for developments in other quantum fluids), it has remained an outstanding challenge to fully map out the in situ velocity field of an atomic superfluid.

\begin{figure*}[t]
    \centering
    \includegraphics[width=\linewidth]{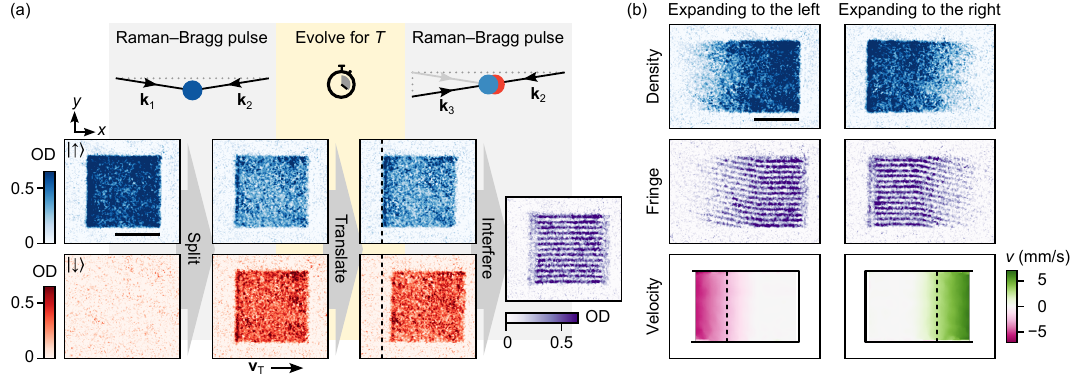}
    \caption{Interferometric velocimetry
    for a 2D atomic superfluid. (a) Experimental protocol. 
   We interfere a cloud prepared in state $\ket{\uparrow}$ (blue) with a slightly displaced copy of itself, using two two-photon Raman-Bragg (RB) pulses. The first pulse splits off a copy of the cloud into the $\ket{\downarrow}$ state (red) and gives it a velocity kick $\mathbf{v}_\text{T} = \hbar(\mathbf{k}_1 - \mathbf{k}_2)/m \approx 25\, {\rm mm}/{\rm s}$ along the $x$ direction; $\mathbf{k}_{1,2}$ are the wave vectors of the two RB laser beams (see cartoon). A time $T$ later (here $T=150~\upmu$s), the second RB pulse, generated using beams with wave vectors $\mathbf{k}_2$ and $\mathbf{k}_3 = -\mathbf{k}_2$, interferes the two copies displaced by $\mathbf{r}_\text{T}$ (here $|\mathbf{r}_\text{T}|\approx 4\,\um$); note that $\mathbf{r}_\text{T} \approx \mathbf{v}_\text{T}T$, but we also take into account the duration ($\approx5\,\upmu$s) of the RB pulses~\cite{SMFootnote}. The resulting interference pattern (purple) shows fringes with a carrier wave vector $\mathbf{q} = \mathbf{k}_1 - \mathbf{k}_3$, with $|\mathbf{q}|\approx2.5\,\um^{-1}$. OD stands for optical density. The local fringe phase reveals the condensate's phase gradient along $x$, and hence $v_x$. In this example, $v_x = 0$, and we observe the straight and uniformly spaced high-contrast fringes expected for a stationary coherent cloud.
    (b) Example measurements. Here we generate 1D Riemann rarefaction waves by removing one (left or right) wall of a square box at $t = 0$. The top row shows the density distributions at $t = 3\,$ms, and the middle row shows the corresponding interference fringes, which bend up (resp. down) for negative (resp. positive) $v_x$. The bottom row shows the extracted $v_x ({\bf r})$ averaged over 11 independent measurements. The solid lines indicate the trap boundaries during the expansion and the dashed lines indicate the box edges removed at $t = 0$. The scale bars in (a) and (b) correspond to 20$\,\um$. 
    }
    \label{fig:2}
\end{figure*}

In this Letter, we present local velocimetry for atomic superfluids in 2D. Using matter-wave interferometry based on Raman--Bragg (RB) beam splitters, we directly measure the local superfluid-phase gradients and quantitatively reconstruct the full 2D velocity field with micrometer-scale resolution.
We demonstrate the capabilities of this technique by measuring the velocity fields in two paradigmatic scenarios: the self-similar dynamics of 1D Riemann rarefaction waves and the quantized circulation around quantum vortices.

Our experiments are based on a homogeneous Bose gas of $\potassium$ atoms in a 2D box trap~\cite{Chomaz:2015, Navon:2021}, as described in~\cite{YZhang:2026}. For different measurements, the gas is confined to the $x-y$ plane by a harmonic trap with frequency $\omega_z/(2\pi)$ ranging from $1.2$ to $\SI{1.7}{\kilo \hertz}$, the linear size of the in-plane box trap ranges from $34$ to $54\,\upmu \rm{m}$, and the 2D atom density $n$ ranges from $34$ to $50~\um^{-2}$. We prepare the gas in the hyperfine ground state $\ket{F,m_F} = \ket{1,-1}$ (labeled in the low-field basis), which we denote $\ket{\uparrow}$, and work in the deeply degenerate regime, where the system is an essentially pure condensate. We work at a magnetic field $B = 41.1$\,G, where the $s$-wave scattering length in our $\ket{\uparrow}$ state is $a_{\uparrow\uparrow}=129 a_0$. The corresponding 2D interaction parameter, $\tilde{g}_{\uparrow\uparrow}= \sqrt{8\pi m \omega_z/\hbar} \, a_{\uparrow\uparrow}$~\cite{Hadzibabic:2011}, ranges from $0.075$ to $0.090$, and the chemical potential, $\mu_{\uparrow} = \hbar^2 \tilde{g}_{\uparrow\uparrow}n/m$,  is always $< 0.7 \,\hbar\omega_z$.  Finally, for our interferometric protocols we employ an auxiliary hyperfine state $\ket{\downarrow} \equiv \ket{1,0}$; at our $B$ field, $a_{\downarrow\downarrow}=-4a_0$ and $a_{\uparrow\downarrow} = -57a_0$, so a trapped mixture of our $\ket{\uparrow}$ and $\ket{\downarrow}$ states is unstable, but this instability is not relevant on the sub-millisecond time scale of our protocols.

\begin{figure*}[t]
    \centering
      \includegraphics[width=1\linewidth]{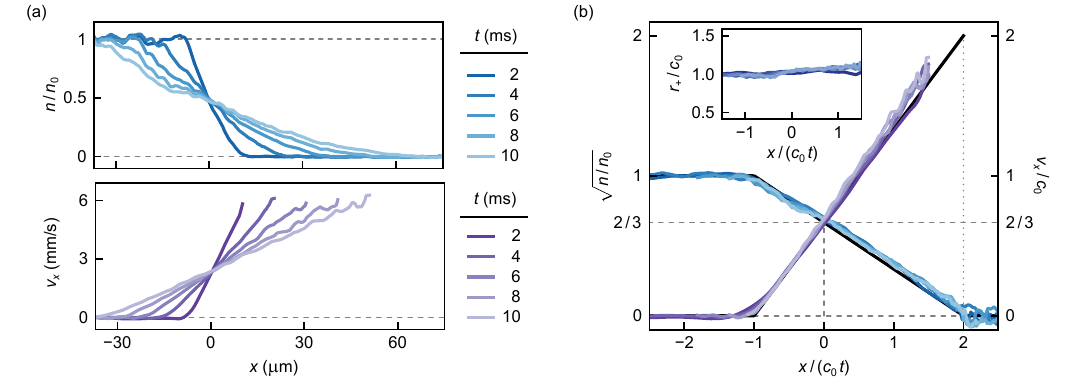}
    \caption{Self-similar Riemann rarefaction flow. 
    (a) Plots of the density and velocity profiles, $n(x, t)$ and $v_x(x, t)$, at different times, $t$, in a 1D Riemann rarefaction wave generated using the same protocol as in Fig.~\ref{fig:2}(b). Here, the condensate expands in the $+x$ direction, and the box wall removed at $t=0$ defines $x = 0$. The condensate density prior to the release is $n_0$, and the corresponding speed of sound is $c_0=3.4(1)~\text{mm}/\text{s}$. 
    (b) The same data scaled by $n_0$ and $c_0$, and plotted versus the similarity variable $\chi = x/(c_0 t)$, collapses onto universal curves. The solid black lines show the analytical predictions from Eq.~(\ref{eq:rarefaction-prediction}). The inset shows the conservation of the Riemann invariant $r_+$ (see the text). Note that, to indicate the noise level around $n=0$ for $x/(c_0t)\gtrsim 2$, wherever the experimentally extracted $n$ is negative, we plot $- \sqrt{|n|/n_0}$. Each curve is based on the mean of approximately $20$ independent measurements, and for the velocity measurements $|\mathbf{r}_{\rm{T}}|\approx2~\um$.
    }
    \label{fig:3}    
\end{figure*}

In Fig.~\ref{fig:2}(a), we present our protocol for measuring one component of the 2D velocity field, here $v_x ({\bf r})$. Using two time-separated Raman--Bragg (RB) two-photon pulses that couple states $\ket{\uparrow}$ and $\ket{\downarrow}$, we interfere our superfluid with a slightly displaced copy of itself, which reveals the component of the local velocity along the displacement direction.

The first RB pulse, generated using two laser beams with wave vectors ${\bf k}_{1}$ and ${\bf k}_{2}$ (see cartoons), splits off a copy of the original condensate (blue) into state $\ket{\downarrow}$ (red) and gives it a velocity kick ${\bf v}_{\rm T}\propto ({\bf k}_{1} - {\bf k}_{2})$ along the $x$ direction. A time $T$ later, we interfere the two clouds using a  $\pi/2$ RB pulse generated using laser beams with wave vectors ${\bf k}_{3}$ and ${\bf k}_{2}$. The resulting fringes have a carrier wave vector $\mathbf{q} = \mathbf{k}_1 - \mathbf{k}_3$, and the atom density in state $\ket{\uparrow}$ is
\begin{equation}
    n_\uparrow(\mathbf{r}) \propto \left|\psi(\mathbf{r}) + \eta  e^{i\phi_\text{RB}} \psi(\mathbf{r} - \mathbf{r}_\text{T}) e^{i\mathbf{q}\cdot\mathbf{r}}\right|^2\,,
    \label{eq:n-up}
\end{equation}
where $\mathbf{r}_\text{T} \approx {\bf v}_{\rm T} T$ is the displacement between the two clouds, $\phi_\text{RB}$ is the global phase between the two RB pulses, and $\eta = 1.86$ reflects the population transfer in the first pulse, which is chosen to balance the local chemical potentials of the two clouds during the translation time $T$ (see~\cite{SMFootnote}).

Previously, similar interferometers~\cite{Hagley:1999b, Hugbart:2005, Clade:2009, Navon:2015,Rabga:2024} have been used to study the coherence properties of atomic clouds by measuring, as a function of $|\mathbf{r}_\text{T}|$, the spatially averaged \textit{contrast} of the interference fringes, as defined by the magnitude of the Fourier component of $n_\uparrow(\mathbf{r})$ at wave vector $\mathbf{q}$.
Here, for small $|\mathbf{r}_\text{T}|$, we extract the local fringe \textit{phase}, which reveals the local velocity in a coherent superfluid. The density $n_\uparrow$ in Eq.~(\ref{eq:n-up}) oscillates as $\cos[\mathbf{q}\cdot\mathbf{r} +\phi_\text{RB} + \phi(\mathbf{r})]$, and the local fringe shift,
\begin{equation}
    \phi(\mathbf{r}) =  \theta(\mathbf{r}-\mathbf{r}_\text{T})- \theta(\mathbf{r})\approx -\mathbf{r}_\text{T}\cdot\nabla\theta(\mathbf{r})\,,
\end{equation}
directly gives 
\begin{equation}
    v_x(\mathbf{r})= -\hbar\phi(\mathbf{r})/(m|\mathbf{r}_\text{T}|)\,.
\end{equation}
Note that in our experiments the global $\phi_{\rm{RB}}$ is not controlled and varies randomly between experimental runs, but can be \emph{a posteriori} calibrated (see~\cite{SMFootnote}).

In our protocol demonstration in Fig.~\ref{fig:2}(a), we start with a stationary equilibrium superfluid, so $v_x=0$, and observe straight and uniformly spaced interference fringes, which provide the baseline for measurements of nonzero $v_x$. 
We show examples of such measurements in Fig.~\ref{fig:2}(b), for rarefaction flows initiated by removing one of the box-trap walls. We show how the interference fringes bend up (resp. down) for negative (resp. positive) $v_x$. To reconstruct $v_x({\bf r})$ from the observed fringes, we use a Fourier-transform method~\cite{Takeda:1982, Takeda:2013} that has a spatial resolution of $\approx\pi/|\mathbf{q}|$; with our $|\mathbf{q}|\approx2.5~\um^{-1}$, we can resolve variations in $v_x$ on a micrometer length scale.

\begin{figure*}[tp]
    \centering
     \includegraphics[width=1\linewidth]{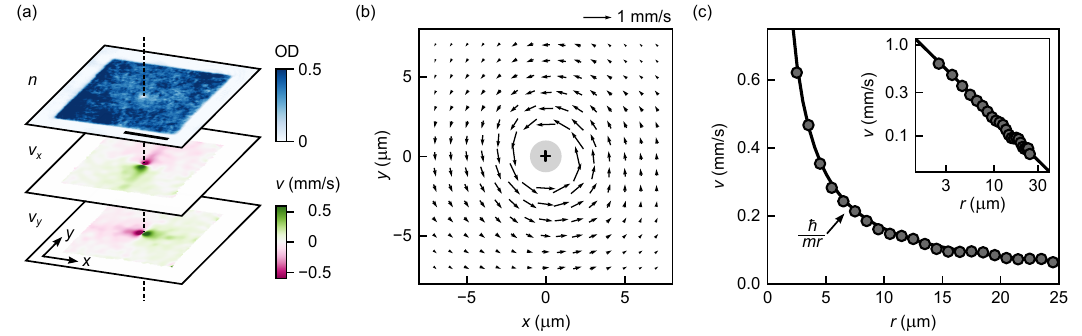}
    \caption{Quantized circulation around a singly-charged vortex. (a) Measured density and velocity fields for a deterministically created vortex at the center of the box ($\mathbf{r} = 0$). See the text for the interferometric protocol for measuring the $y$ component of the velocity, $v_y$. (b) Quiver plot showing the reconstructed 2D velocity field. (c) Plot of the scalar speed $v=|\mathbf{v}|$ versus the radial distance, $r$, from the vortex; the inset shows a log-log plot of the same data. The solid lines show the theoretical prediction $v = \hbar/(mr)$. For both $v_x$ and $v_y$ measurements, $|\mathbf{r}_{\rm{T}}|\approx2~\um$, and the data shown is based on the mean of approximately 300 independent measurements. The scale bar in (a) corresponds to 20$\,\um$.
    }
    \label{fig:4}    
\end{figure*}

As the first application of our velocimetry, we study the evolution of the $v_x({\bf r})$ field such as shown in Fig.~\ref{fig:2}(b). For our initially uniform superfluid with a sharp edge, this expansion into vacuum realizes a 1D Riemann rarefaction flow, an archetypal nonlinear hydrodynamic flow that  admits analytical solutions~\cite{Riemann:1860, Landau:1959}. 

For fluids with a scale-invariant polytropic equation of state, $P\propto n^\gamma$, where $P$ is the hydrodynamic pressure and $\gamma$ the polytropic exponent, theory predicts self-similar evolution of both density and velocity fields~\cite{Riemann:1860, Landau:1959}. The self-similar density dynamics were recently observed for strongly interacting 3D Fermi gases, for which $\gamma=5/3$~\cite{Wolf:2026}, and related expansion dynamics have also been explored for harmonically trapped Bose gases~\cite{Hoefer:2006, Sharan:2025}. Here, we observe self-similar evolution of both density and velocity fields for our (initially) homogeneous Bose gas, with $P =g n^2/2$, where $g=\hbar^2 \tilde{g}_{\uparrow\uparrow}/m$, arising from interatomic interactions~\cite{Pitaevskii:2016}.

In the hydrodynamic description, the 1D equations of motion for a weakly interacting condensate are~\cite{Pitaevskii:2016}
\begin{equation}
    \frac{\partial n}{\partial t} + \frac{\partial}{\partial x}(n v_x) = 0 ~~ \text{and} ~~
    \frac{\partial v_x}{\partial t} + v_x \frac{\partial v_x}{\partial x} = - \frac{1}{n m} \frac{\partial P}{\partial x}\,.
    \label{eq:hydro-EoM}
\end{equation}
Under these equations, when a condensate expands into vacuum from an initial density discontinuity, there is no intrinsic spatio-temporal scale apart from the initial (in-trap) speed of sound $c_0$~\cite{Landau:1959, Wolf:2026}, given by $c_0= \sqrt{gn_0/m}$, where $n_0$ is the initial density. The density $n(x,t)$ and velocity $v_x(x,t)$ fields 
can then depend on $x$ and $t$ only through the similarity variable $\chi= x/ (c_0 t)$, and must evolve self-similarly. Writing $n(x, t) = n_0 \tilde{n}(\chi)$,  $v_x(x, t) = c_0 \tilde{v}_x (\chi)$, and $P = mc_0^2 n_0 \tilde{n}^2(\chi)/2$, Eq.~(\ref{eq:hydro-EoM}) is solved by 
\begin{equation}
    \sqrt{\tilde{n}} = (2  - \chi)/3 ~~~ \text{and} ~~~ \tilde{v}_x = 2(\chi +1)/3
    \label{eq:rarefaction-prediction}
\end{equation}
inside the rarefaction wave, $-1\leq\chi\leq 2$, and constant $\tilde{n}$ and $\tilde{v}_x$ outside. Note that $r_+= c + v_x/2 =c_0$, where $c = c_0\sqrt{\tilde{n}}$ is the local speed of sound, is a Riemann invariant~\cite{Landau:1959, Whitham:1974}, conserved across the rarefaction wave~\cite{Ivanov:2019, Sharan:2025}. 

In Fig.~\ref{fig:3}, we experimentally confirm these analytical predictions. Here, the right wall of the box, which defines $x=0$, is removed at $t=0$, and the initial speed of sound, measured as in Ref.~\cite{Ville:2018}, is $c_0=3.4(1)~\text{mm}/\text{s}$. In Fig.~\ref{fig:3}(a), we plot $n(x, t)$ and $v_x(x, t)$ for different $t$, and in Fig.~\ref{fig:3}(b) we show that they collapse onto universal curves when scaled, respectively, by $n_0$ and $c_0$, and plotted versus $\chi$. These collapsed curves agree with the theoretical predictions in Eq.~(\ref{eq:rarefaction-prediction}), shown by the solid black lines.
We note that Eq.~(\ref{eq:hydro-EoM}) neglects the quantum pressure~\cite{Pitaevskii:2016}, relevant on length scales shorter than the healing length $\xi=\hbar/\sqrt{2m g n_0}\approx0.35~\um$. Consequently, the prediction in Eq.~(\ref{eq:rarefaction-prediction}) is valid only for $t \gg \xi/c_0\approx0.1~$ms, which is well satisfied for all $t$ in our data.
Also note that theoretically $v_x$ is not defined for $x/(c_0 t) > 2$, where $n=0$. Experimentally, we can measure $v_x$  up to $x\approx1.5\,c_0 t$, which corresponds to $n\approx 0.03\,n_0$.

As the second application of our velocimetry, we study the 2D flow around quantized vortices, now separately measuring both $v_x({\bf r})$ and  $v_y({\bf r})$.
Singly charged vortices with $\pm2\pi$ phase winding are elementary topological defects in 2D superfluids that carry quantized angular momenta. They play a central role in, for example, the Berezinskii-Kosterlitz-Thouless transition~\cite{Berezinskii:1971, Kosterlitz:1973, Hadzibabic:2006} and 2D superfluid turbulence~\cite{ Bradley:2026}.
Here, in Fig.~\ref{fig:4}, we deterministically create a single $+1$ charged vortex at the center of our box using the chopstick method~\cite{Samson:2016}.

We measure $v_x(\mathbf{r})$ using the same method as in Fig.~\ref{fig:2}(a). To measure $v_y(\mathbf{r})$, one could introduce another set of three laser beams, rotated by $90^\circ$ relative to those in Fig.~\ref{fig:2}(a). However, here we can also measure $v_y$ more simply, using two of the three beams already set up for the $v_x$ measurements.  Specifically, we now split off a copy of the cloud into the $\ket{\downarrow}$ state using an RB pulse generated with the two beams with wave vectors $\mathbf{k}_1$ and $\mathbf{k}_3$. This copy then moves at $\hbar |\mathbf{q}|/m \approx 4\, {\rm mm}/{\rm s} $ along the $y$ direction. Then, after the evolution time $T$, we interfere the two clouds using a $\pi/2$ radio-frequency pulse, which carries no momentum. In this protocol, the resulting matter-wave fringes have the same carrier wave vector $\mathbf{q}$ as before, but their phase now reveals $v_y(\mathbf{r})$~\footnote{For our parameters, this simpler protocol suffices because $|{\bf q}|$ is both large enough for the two clouds to be displaced by $2\,\upmu{\rm m}$ in $0.5\,$ms (shorter than any other dynamics in the system) and small enough for us to resolve the interference fringes. However, in general, the approach in Fig.~\ref{fig:2}(a) is more versatile, allowing for larger $|{\bf v}_{\rm T}|$ and independently tunable $|{\bf q}|$.}.
Note that due to the slight inhomogeneity of our RB beams, for stationary clouds we observe a static pattern of an apparent velocity field with $|\mathbf{v}| < 0.1 \,\text{mm/s}$. These systematic offsets are insignificant for the measurements in Fig.~\ref{fig:3}, but here they are relevant and we subtract them from the data.

\begin{figure}[h!]
\centering
     \includegraphics[width=1\columnwidth]{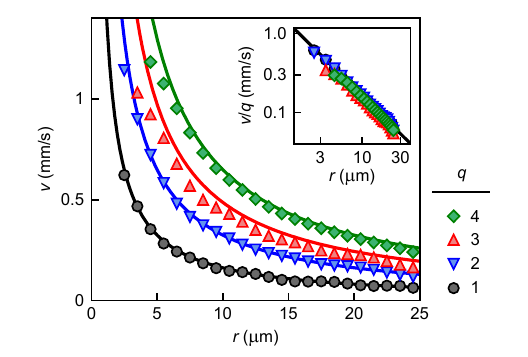}
    \caption{ Circulation around vortices with different charges $q$. By merging multiple singly charged vortices, we deterministically create,  at the cloud center, metastable multiply charged ones. Here, as in Fig.~\ref{fig:4}, we plot the speed $v=|\mathbf{v}|$ versus the radial distance, $r$, from the vortex. The solid lines show the theoretical predictions $v = q\hbar/(mr)$. }
    \label{fig:S3}   
\end{figure}

In Fig.~\ref{fig:4}(a), we show separately measured $n ({\bf r})$, $v_x({\bf r})$, and $v_y({\bf r})$ for a single vortex at $\mathbf{r} = 0$, and in Fig.~\ref{fig:4}(b) we show the quiver plot of the reconstructed 2D velocity field  ${\bf v}({\bf r}) = (v_x ({\bf r}), v_y ({\bf r}))$, which qualitatively shows the expected circulation around a vortex. 
Quantitatively, in Fig.~\ref{fig:4}(c), we show that the azimuthally averaged speed $v = |\mathbf{v}|$ agrees with the theoretical prediction, $v = \hbar/(mr)$, for a single quantum of circulation. 
We have also verified the quantization of the circulation and the irrotationality of the flow by line-integrating ${\bf v}({\bf r})$ around different contours; see Fig.~\ref{fig:S2} in End Matter. This quantized circulation provides an elementary building block for more complex superfluid flows; see also End Matter (Figs.~\ref{fig:S3} and \ref{fig:S4}) for similar measurements of velocity fields around multiply charged vortices and pairs of vortices with either the same or opposite charges.

To conclude, we have demonstrated an interferometric technique that gives direct access to the in situ velocity field in planar atomic superfluids, which complements the traditional probes based on atom-density measurements and offers a new window on superfluid hydrodynamics. Here, we used it to study the classic problem of Riemann rarefaction waves and to directly visualize the quantized irrotational flow around quantum vortices. In the future, the same technique, combined with the flexibility afforded by optical box traps~\cite{Navon:2021}, could be applied to studies of, for example, superflows around obstacles and through channels between reservoirs~\cite{Chien:2015}, or more generally through various atomtronic circuits~\cite{Amico:2021}. Moreover, measurements of velocity correlations could open many further opportunities, for example for studies of vortex interactions~\cite{Kwon:2021} and superfluid turbulence~\cite{Bradley:2026}.

We thank Martin Zwierlein and Nigel Cooper for useful discussions.
This work was supported by EPSRC [Grant No.~EP/Y01510X/1], ERC [UniFlat], and STFC [Grants No.~ST/T006056/1 and No.~ST/Y004469/1]. L.D. was supported by the French government through the Program `Investissements d’avenir' (I-SITE ULNE / ANR-16-IDEX-0004 ULNE), managed by the National Research Agency. C.E. acknowledges support from the Royal Society (URF/R1/261565). G.C. was supported by an NSERC Postdoctoral Fellowship.

\onecolumngrid
\clearpage
\twocolumngrid

\onecolumngrid
\begin{center}
   {\large\bfseries End Matter}
\end{center}

\twocolumngrid

\begin{figure}[t]
\centering
    \includegraphics[width=1\columnwidth]{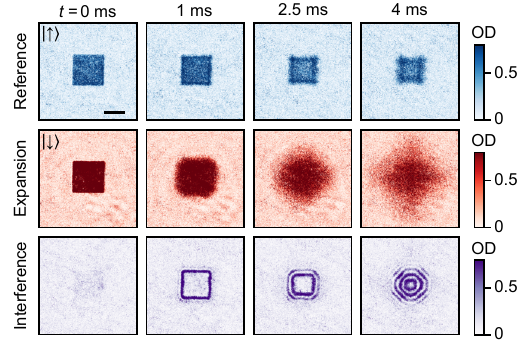}
    \caption{
    Interferometry using a quasi-uniform phase reference. Here, a superfluid comprising two spin states is released from a 2D box trap at $t = 0$; the black scale bar corresponds to $20\,\upmu$m. One state ($\ket{\downarrow}$, red) expands rapidly, while the other one ($\ket{\uparrow}$, blue), with much weaker interatomic interactions, evolves much less and provides an approximately uniform phase reference.
    Interfering the two states, using a RF pulse, results in fringes (purple) that primarily reveal the phase accumulation in the expanding component. 
    }
    \label{fig:S1}   
\end{figure}

\textit{Interferometry with a phase reference} -- In Fig.~\ref{fig:S1}, we qualitatively illustrate how superflow can be interferometrically observed using a separate condensate as a phase reference. Note that here we work at a different magnetic field than in the main paper, $B = 58.1$\,G, where $a_{\uparrow\uparrow}= 31a_0, a_{\downarrow\downarrow}=220a_0$, and $a_{\uparrow\downarrow} = -53a_0$, and that for this illustration the roles of our two hyperfine states are swapped -- `the main system' is in state $\ket{\downarrow}$, and state $\ket{\uparrow}$ is used as an approximate phase reference.

We prepare a stationary superfluid in state $\ket{\uparrow}$, then use a radio-frequency (RF) pulse to transfer $\approx80\%$ of the atom population to state $\ket{\downarrow}$ at time $t=0$ and immediately turn off the box trap. The more strongly interacting $\ket{\downarrow}$ cloud (red) expands significantly within a few milliseconds, whereas, on the same time scale, the $\ket{\uparrow}$ state (blue) evolves much less. 
At different times $t$, we interfere the two spin states using a second RF pulse. This RF pulse is chosen such that, in the region where the two clouds overlap, it transfers essentially all the population out of the $\ket{\uparrow}$ state if the clouds still have the same local phase. Finally, we image the $\ket{\uparrow}$ state, now shown in purple in the bottom row of Fig.~\ref{fig:S1}. The appearance of any local population in this state reveals the accumulation of a phase difference between the two states, which primarily arises due to the expansion of the $\ket{\downarrow}$ component, while the $\ket{\uparrow}$ cloud serves as an approximately uniform phase reference.

The interferograms in the bottom row of Fig.~\ref{fig:S1} show some key qualitative features of the expansion of the $\ket{\downarrow}$ cloud from the 2D box trap. The velocity field first develops at the edges of the box, and the rarefaction waves then propagate inwards to the center of the box. Compared to the 1D rarefaction flow in Fig.~\ref{fig:3}, here the interaction of the waves from the four box edges leads to intricate flow profiles at later times.

While this illustration is only qualitative, in principle, for $|a_{\uparrow\uparrow}|, |a_{\uparrow\downarrow}| \ll a_{\downarrow\downarrow}$, such interferograms could be used to quantitatively reconstruct the velocity field of the $\ket{\downarrow}$ component in the area where the two clouds overlap, by differentiating their relative phase. Note, however, that these interference patterns alone do not unambiguously give the phase field. In particular, one cannot identify the flow direction without additional information, such as the time-resolved density images in the middle row of Fig.~\ref{fig:S1}, because an interferogram with density modulation $\propto\cos[\phi(\mathbf{r})]$ is invariant under $\phi(\mathbf{r})\xrightarrow[]{}-\phi(\mathbf{r})$. In the method presented in the main paper, the injection of a carrier fringe with wave vector $\mathbf{q}$ resolves this ambiguity since $\cos[\mathbf{q}\cdot\mathbf{r}+\phi(\mathbf{r})]\neq\cos[\mathbf{q}\cdot\mathbf{r}-\phi(\mathbf{r})]$ for nonzero $\phi(\mathbf{r})$.

\onecolumngrid
\vspace{2.5\baselineskip}
\noindent
\begin{minipage}{\textwidth}
    \centering
     \includegraphics[width=1\linewidth]{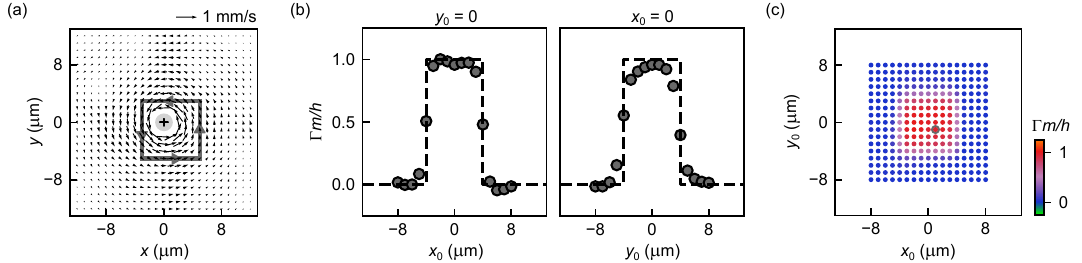}
    \captionof{figure}{
 Quantized circulation and irrotationality of the superflow. (a) Quiver plot of the velocity field for a singly charged vortex at the center of the box, as in Fig.~\ref{fig:4}(b). We extract the circulation $\Gamma$ given by the line integral of the velocity field along an \(\SI{8}{\micro \metre} \times \SI{8}{\micro \metre}\) square contour (gray) centered at ($x_0$, $y_0$). (b) Normalized circulation \(\Gamma m / h\) as a function of $x_0$ for fixed $y_0 =0$ and vice versa. The dashed lines show the theoretical predictions. (c) 2D map of the normalized circulation. We find $\Gamma \approx h/m$ for all $(x_0, y_0)$ such that the contour encloses the vortex, and $\Gamma \approx 0$ otherwise. The gray circle indicates the position (\(x_0=\SI{1}{\micro \meter}, y_0=\SI{-1}{\micro \meter}\)) of the contour shown in (a).
}
    \label{fig:S2}   
\end{minipage}
\onecolumngrid
\clearpage
\twocolumngrid

\textit{Irrotationality of the superflow} -- 
In Fig.~\ref{fig:S2}, we show an additional confirmation of the quantization of the velocity circulation and a verification of the irrotationality of the superflow. Here, for a singly charged vortex in the center of the box, $x=y=0$ (as in Fig.~\ref{fig:4}), we calculate the circulation
\begin{equation}
\Gamma = \oint_{C(x_0, y_0)} \mathbf{v} \cdot \mathrm{d}\mathbf{l} \, 
\end{equation}
for square contours $C$ centered at different positions $(x_0,y_0)$.
As expected for quantized superflow, we observe $\Gamma \approx h/m$ as long as the contour encloses the vortex, and $\Gamma \approx 0$ otherwise.

\textit{Additional vortex configurations} -- Here we show measurements of the velocity fields for additional vortex configurations. In Fig.~\ref{fig:S3}, we show $v(r)$ for multiply charged vortices. In Fig.~\ref{fig:S4}, we show the velocity fields around pairs of singly charged vortices that have either the same or opposite charges.

\onecolumngrid
\vspace{2.5\baselineskip}
\noindent
\begin{minipage}{\textwidth}
    \centering
     \includegraphics[width=1\linewidth]{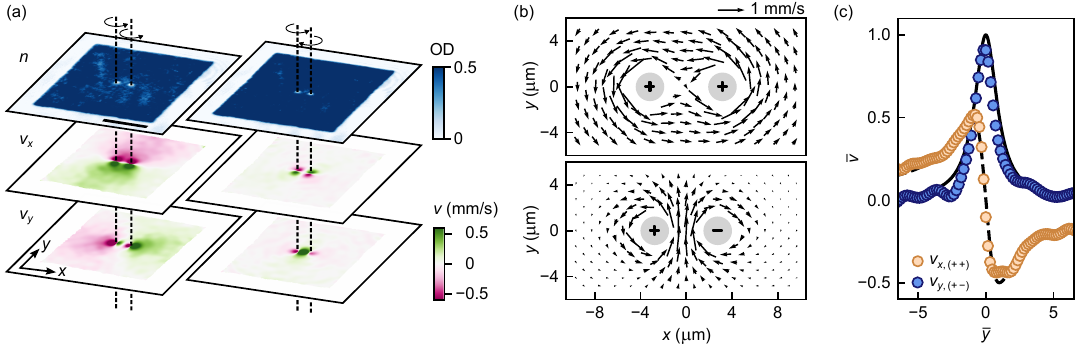}
    \captionof{figure}{
    Superflows around vortex pairs. 
    (a)~Measured density and velocity fields for deterministically prepared pairs of vortices of either the same sign (left) or opposite signs (right).
(b)~The corresponding quiver plots of the 2D velocity fields.  
The vortex core positions (gray circles), inferred from the density depletions, are separated by $2d$ along $x$, with $d=3.15\,\upmu\rm{m}$ for the same-sign vortices ($+,+$) and $d=2.75\,\upmu\rm{m}$ for the opposite-signs ones ($+,-$) .
(c)~Scaled velocity $\bar{v} = dmv/(2\hbar)$ versus $\bar{y} = y/d$, along the symmetry axis of the vortex pairs.  
The lines show the theoretical predictions $\bar{v} = -\bar{y}(1+\bar{y}^2)^{-1}$ for $v_x$ in the ($+,+$) case (dashed) and $\bar{v} = (1+\bar{y}^2)^{-1}$ for $v_y$ in the ($+,-$) case (solid).
For all measurements $|\mathbf{r}_{\rm{T}}|\approx2~\um$, and the data was averaged over approximately 300 independent measurements. The scale bar in (a) corresponds to 20$\,\um$.
}
    \label{fig:S4}   
\end{minipage}
\onecolumngrid
\clearpage
\twocolumngrid

\section{Supplemental material}

\subsection{Interaction effects during the interferometry sequence}

Interatomic interactions can lead to additional phase shifts in our interferometry sequence and contaminate the velocity measurement. 
For a two-component condensate with wavefunction $(\psi_\uparrow(\mathbf{r}), \psi_\downarrow(\mathbf{r}))^\text{T}$ and a slowly varying density profile, the two local chemical potentials are
\begin{equation}
\begin{split}
    \mu_\uparrow &= g_{\uparrow\uparrow} n_\uparrow + g_{\uparrow\downarrow}n_\downarrow \, ,\\
    \mu_\downarrow &= g_{\downarrow\downarrow} n_\downarrow + g_{\uparrow\downarrow}n_\uparrow\,,
\end{split}
\end{equation}
with $n_{\uparrow, \downarrow}=|\psi_{\uparrow, \downarrow}(\mathbf{r})|^2$ and 
$g_{ij}=(\hbar^2/m) \sqrt{8\pi m\omega_z/\hbar}\,a_{ij}$ for $i,j\in\{\uparrow, \downarrow\}$.
Any difference between $\mu_{\uparrow}$ and $\mu_{\downarrow}$ leads to accumulation of the local relative phase $\varphi$ between the two components, at a rate
\begin{equation}
    \hbar \dot{\varphi} = \mu_\uparrow - \mu_\downarrow = \frac{1}{2}(g_{\uparrow\uparrow} + g_{\downarrow\downarrow}-2g_{\uparrow\downarrow})(Z - Z_0)\,n \,,
\end{equation}
with $Z = (n_\uparrow - n_\downarrow)/(n_\uparrow + n_\downarrow)$ , $n = n_\uparrow + n_\downarrow$, and $Z_0=-(g_{\uparrow\uparrow}-g_{\downarrow\downarrow})/(g_{\uparrow\uparrow}+g_{\downarrow\downarrow}-2g_{\uparrow\downarrow})$.

To suppress these interaction effects, we adjust the first RB pulse such that $Z=Z_0$ after this pulse. This ensures $\mu_\uparrow = \mu_\downarrow$  during the evolution time $T$ for any local density $n$. For our measurements at $B = 41.1~$G, $Z_0=-0.55$, so we use a $0.69\,\pi$ pulse in the first step of our sequence.

\subsection{Calibrating $\phi_{\rm{RB}}$}

In our study of the Riemann waves in Figs.~\ref{fig:2} and \ref{fig:3}, we fix $\phi_{\rm{RB}}$ by imposing that $v_x = 0$ at the box edge opposite to the removed wall.
For the studies of vortices (Figs.~\ref{fig:4}, \ref{fig:S2}, \ref{fig:S3}, \ref{fig:S4}), we fix $\phi_{\rm{RB}}$ by imposing the condition that the spatially averaged velocity vanishes ($\langle v_x\rangle = \langle v_y\rangle = 0$). 
More generally, one could always make two side-by-side box traps, of which one contains a stationary superfluid. Our interferometry protocol would simultaneously act on both clouds, and the straight uniformly spaced fringes in the stationary one [as in Fig.~\ref{fig:2}(a)] would provide the $\phi_{\rm{RB}}$ calibration in each experimental shot.

\subsection{Effect of the finite duration of spin-transfer pulses}

In the limit of instantaneous spin transfers, the relative translation $\mathbf{r}_\text{T}$ in Eq.~(\ref{eq:n-up}) is simply $\mathbf{v}_\text{T} T$. Here, we calculate the small ($\approx10\%$) corrections to $\mathbf{r}_\text{T}$ due to the finite duration of the spin-transfer pulses.
We first analyze the motion of atoms during different spin-transfer pulses, and then calculate their combined effect on $\mathbf{r}_\text{T}$.  Note that here we ignore interaction effects and the dispersion of the atomic wave packet during the measurement sequence.

\subsubsection{RF and RB pulses}

An RF pulse does not modify the momentum of the atoms, so a moving component just keeps moving during the pulse. For example, for atoms initially in the $\ket{\downarrow}$ component with wavefunction $\psi(\mathbf{r})e^{i\mathbf{K}\cdot\mathbf{r}}$, a $\pi/2$ RF pulse of duration $\tau$ coupling the $\ket{\uparrow}$ and $\ket{\downarrow}$ components leads to
\begin{equation}
    \psi(\mathbf{r})e^{i\mathbf{K}\cdot\mathbf{r}}\ket{\downarrow} \xrightarrow{} \psi(\mathbf{r} - \mathbf{v}_\mathbf{K} \tau) e^{i\mathbf{K}\cdot\mathbf{r}}(i\ket{\uparrow} + \ket{\downarrow})/\sqrt{2}\,.
\end{equation}
Here, $\psi(\mathbf{r})$ is a generic wavefunction that varies slowly in space, and $e^{i\mathbf{K}\cdot\mathbf{r}}$ encodes the center-of-mass velocity $\mathbf{v}_\mathbf{K}=\hbar\mathbf{K}/m$ of the atomic wave packet. 

An RB pulse, on the other hand, imparts momentum, and the velocity of the atoms changes during the pulse.
Consider an RB pulse that couples plane-wave modes $\ket{\mathbf{k}, \uparrow}$ and $\ket{\mathbf{k}+\mathbf{K}, \downarrow}$, where $\ket{\mathbf{k}, \sigma}$ denotes a plane-wave state in component $\sigma\in\{\uparrow, \downarrow\}$. The coupling Hamiltonian in the rotating-wave approximation is
\begin{equation}
    H_\mathbf{k}  =  \begin{pmatrix}
        \varepsilon_\mathbf{k} &-\hbar\Omega/2\\
        -\hbar\Omega/2 & \varepsilon_\mathbf{k} + \hbar^2\mathbf{k}\cdot\mathbf{K}/m + \varepsilon_\mathbf{K}+\hbar \delta
    \end{pmatrix}\,.
\end{equation}
Here, $\varepsilon_\mathbf{k} = \hbar^2|\mathbf{k}|^2/(2m)$ is the kinetic energy for wave vector $\mathbf{k}$, $\Omega$ is the Rabi frequency, and $\delta$ is the two-photon detuning from the atomic transition frequency. 
We always set $\hbar\delta=-\varepsilon_\mathbf{K}$, so the coupling is near-resonant for $|\mathbf{k}|\approx0$. The evolution during a square RB pulse of duration $\tau$ is described by the matrix $U_\mathbf{k} = e^{-iH_\mathbf{k} \tau/\hbar}$, which is approximately
\begin{equation}
\begin{split}
   &U_\mathbf{k}= 
   e^{-i \mathbf{k}\cdot \mathbf{L}}  \\
   &\times
   \begin{pmatrix}
       \cos({\vartheta}/{2})\left[1 + i\mathbf{k}\cdot\mathbf{L}'\right] & i\sin({\vartheta}/{2})\\
       i\sin({\vartheta}/{2}) & \cos({\vartheta}/{2})\left[1 - i\mathbf{k}\cdot\mathbf{L}'\right]
   \end{pmatrix}\,.
   \label{eq:U-matrix}
\end{split}
\end{equation}
Here, $\mathbf{L} =\mathbf{v}_\mathbf{K} \tau/2$ and $\mathbf{L}'=\mathbf{v}_\mathbf{K} \tau\tan(\vartheta/2)/\vartheta$, where  $\mathbf{v}_\mathbf{K}=\hbar\mathbf{K}/m$ and $\vartheta=\Omega \tau$. 
Note that we retain terms only up to order $\Omega^{-1}$, since $\Omega\approx2\pi\times50~\mathrm{kHz}$ is the largest frequency (energy) scale in our system. We also neglect phases  $\sim\varepsilon_{\mathbf{k}}\tau/\hbar\ll 1$, which cause only a weak dispersion of the wave packets.
For $\mathbf{k}\cdot\mathbf{L}'\ll1$, which is always satisfied for our experimental parameters as long as $\vartheta$ is not very close to $\pi$, we can further approximate $U_\mathbf{k}$ as
\begin{equation}
    U_\mathbf{k} = e^{-i \mathbf{k}\cdot \mathbf{L}}  \begin{pmatrix}
       e^{i\mathbf{k}\cdot \mathbf{L}'}\cos({\vartheta}/{2}) & i\sin({\vartheta}/{2})\\
       i\sin({\vartheta}/{2}) & e^{-i\mathbf{k}\cdot \mathbf{L}'}\cos({\vartheta}/{2})
   \end{pmatrix}\,.
\end{equation}

By applying $U_{\mathbf{k}}$ on each plane-wave mode in a wave packet, one can show that the RB pulse leads to
\begin{equation}
\begin{split}
    \psi(\mathbf{r})\ket{\uparrow} \xrightarrow{} & \cos(\vartheta/2)\psi(\mathbf{r} - \mathbf{L} + \mathbf{L}')\ket{\uparrow} \\ &+ i\sin(\vartheta/2)\psi(\mathbf{r} - \mathbf{L}) e ^{i\mathbf{K}\cdot\mathbf{r}}\ket{\downarrow}\,,
\end{split}
\end{equation}
and similarly,
\begin{equation}
\begin{split}
    \psi(\mathbf{r})e^{i\mathbf{K}\cdot\mathbf{r}}\ket{\downarrow} \xrightarrow{} &\cos(\vartheta/2)\psi(\mathbf{r} - \mathbf{L} - \mathbf{L}')e^{i\mathbf{K}\cdot\mathbf{r}}\ket{\downarrow}\\
    &+ i \sin(\vartheta/2)\psi(\mathbf{r} - \mathbf{L})\ket{\uparrow}\,.
\end{split}
\end{equation}
Note that $\mathbf{L}'\approx\mathbf{L}$ for $\vartheta \ll 1$. 

\subsubsection{$v_y$ \textrm{measurements}}
In our $v_y$ measurements, we start with atoms in state $\ket{\psi} = \psi(\mathbf{r}) \ket{\uparrow}$, and the RB pulse with duration $\tau_\text{I}$ and recoil momentum $\hbar\mathbf{q} = \hbar(\mathbf{k}_1 - \mathbf{k}_3)$ [see Fig.~\ref{fig:2}(a)] leads to
\begin{equation}
\begin{split}
    \ket{\psi} = & \cos(\vartheta_\text{I}/2) \psi(\mathbf{r} - \mathbf{L}_\text{I} + \mathbf{L}_\text{I}')\ket{\uparrow}\\  & + i \sin(\vartheta_\text{I}/2) \psi(\mathbf{r} -\mathbf{L}_\text{I}) e^{i\mathbf{q}\cdot\mathbf{r}}\ket{\downarrow}\,.
\end{split}
\end{equation}
Here, $\mathbf{L}_\text{I} = \mathbf{v}_\text{T} \tau_\text{I}/2$, $\mathbf{L}_\text{I}' = \tan(\vartheta_\text{I}/2)/\vartheta_\text{I}\cdot \mathbf{v}_\text{T} \tau_\text{I}$, $\mathbf{v}_\text{T} = \hbar\mathbf{q}/m$.
Following the translation time $T$, we have
\begin{equation}
\begin{split}
    \ket{\psi} = & \cos(\vartheta_\text{I}/2) \psi(\mathbf{r} - \mathbf{L}_\text{I} + \mathbf{L}_\text{I}')\ket{\uparrow}\\  & + i \sin(\vartheta_\text{I}/2) \psi(\mathbf{r} -\mathbf{L}_\text{I} - \mathbf{v}_\text{T} T) e^{i\mathbf{q}\cdot\mathbf{r}}\ket{\downarrow}\,,
\end{split}
\end{equation}
and the projection of $\ket{\psi}$ onto the $\ket{\uparrow}$ component after the $\pi/2$ RF pulse of duration $\tau_\text{II}$ is
\begin{equation}
\begin{split}
    \braket{\uparrow|\psi} \propto & \cos(\vartheta_\text{I}/2) \psi(\mathbf{r} - \mathbf{L}_\text{I} + \mathbf{L}_\text{I}') \\
    &  - \sin(\vartheta_\text{I}/2) \psi[\mathbf{r}  -\mathbf{L}_\text{I} - \mathbf{v}_\text{T} (T + \tau_\text{II})] e^{i\mathbf{q}\cdot\mathbf{r}}\,.
\end{split}
\label{eq:v-y-up-wavefunc}
\end{equation}
To image $n_\uparrow$, we use a further RF $\pi$-rotation, with pulse time $\tau_\text{III}$, to transfer atoms in state $\ket{\uparrow}$ to the imaging state $\ket{F,m_F}=\ket{2,-2}$; during this pulse the wave packet described by the second term in Eq.~(\ref{eq:v-y-up-wavefunc}) continues translating along $\mathbf{q}$. At the time of imaging, the total relative displacement is $\mathbf{r}_\text{T} = \mathbf{v}_\text{T} T_\text{eff}$
with
\begin{equation}
    T_\text{eff} = T + \tan(\vartheta_\text{I}/2)\tau_\text{I}/\vartheta_\text{I} + \tau_\text{II} + \tau_\text{III}\,.
\end{equation}
In our experiments, $T=500~\upmu$s, $\vartheta_\textrm{I} =0.69\, \pi$, $\tau_\text{I}\approx 5~\upmu$s, $\tau_\text{II}\approx 25~\upmu$s, and $\tau_\text{III}\approx 35~\upmu$s. These parameters result in a $\approx13\%$ difference between $T_\text{eff}$ and $T$.

\subsubsection{$v_x$ \textrm{measurements}}
In our $v_x$ measurements, just prior to the second RB pulse, the atomic wavefunction is
\begin{equation}
\begin{split}
    \ket{\psi} = & \cos(\vartheta_\text{I}/2) \psi(\mathbf{r} - \mathbf{L}_\text{I} + \mathbf{L}_\text{I}')\ket{\uparrow}\\  & + i \sin(\vartheta_\text{I}/2) \psi(\mathbf{r} -\mathbf{L}_\text{I} - \mathbf{v}_\text{T} T) e^{i\mathbf{Q}\cdot\mathbf{r}}\ket{\downarrow}\,.
\end{split}
\end{equation}
Here, $\mathbf{Q} = \mathbf{k}_1 - \mathbf{k}_2$ and $\mathbf{v}_\text{T} = \hbar\mathbf{Q}/m$. After the second RB pulse, with rotation angle $\vartheta_\text{II} = \pi/2$, duration $\tau_\text{II}$, and recoil momentum $\hbar\mathbf{Q}' = \hbar(\mathbf{k}_3 - \mathbf{k}_2)$, we have
\begin{equation}
\begin{split}
    \braket{\uparrow|\psi} \propto & \cos(\vartheta_\text{I}/2) \psi(\mathbf{r} - \mathbf{L}_\text{I} + \mathbf{L}_\text{I}'- \mathbf{L}_\text{II} + \mathbf{L}_\text{II}') \\
    &  - \sin(\vartheta_\text{I}/2) \psi[\mathbf{r}  -\mathbf{L}_\text{I} - \mathbf{v}_\text{T}T - \mathbf{L}_\text{II}] e^{i\mathbf{q}\cdot\mathbf{r}}\,.
\end{split}
\label{eq:v-x-up-wavefunc}
\end{equation}
Here, $\mathbf{L}_\text{II} \approx \mathbf{v}_\text{T} \tau_\text{II}/2$ and $\mathbf{L}_\text{II}' \approx \tan(\vartheta_\text{II}/2)/\vartheta_\text{II}\cdot \mathbf{v}_\text{T} \tau_\text{II}$. Note that in the expressions for $\mathbf{L}_\text{II}$ and $\mathbf{L}_\text{II}'$, we have ignored the small difference between $\mathbf{Q}$ and $\mathbf{Q}'$. Finally, as for the $v_y$ measurements, we image $n_\uparrow$ after a $\pi$-pulse transfer to $\ket{2, -2}$, and the total relative displacement at the time of imaging is
\begin{equation}
    \mathbf{r}_\text{T} = \mathbf{v}_\text{T} T_\text{eff} + \hbar\mathbf{q}/m \cdot \tau_\text{III}\,,
\end{equation}
with
\begin{equation}
    T_\text{eff} = T + \tan(\vartheta_\text{I}/2)\tau_\text{I}/\vartheta_\text{I} + \tan(\vartheta_\text{II}/2)\tau_\text{II}/\vartheta_\text{II}\,.
\end{equation}
In our experiments, $\tau_\text{I}\approx\tau_\text{II}\approx 5~\upmu$s, $\vartheta_\text{I}=0.69\pi$, and $\vartheta_\text{II} = \pi/2$. For our $v_x$ measurements in Figs.~\ref{fig:3} and \ref{fig:4}, $T=80~\upmu$s, and the finite RB pulse durations lead to a $\approx10\%$ difference between $T_\text{eff}$ and $T$. For $|\mathbf{v}_\text{T} T_\text{eff}|\approx 2~\um$, the small relative translation by $|\hbar\mathbf{q}/m \cdot \tau_\text{III}|\approx 0.14~\um$ along the $y$ axis during the RF transfer for the imaging does not appreciably modify $|\mathbf{r}_\text{T}|$ and only leads to a small ($\approx4^\circ$) tilt of $\mathbf{r}_\text{T}$ away from the $x$ axis. We ignore the latter effect in our data analysis.

\clearpage

\end{document}